\documentclass[twocolumn, showpacs, aps, prl, superscriptaddress,
floatfix]{revtex4}
\usepackage{graphicx}
\usepackage{amsmath}

\newcommand{\bo}{\bar{1}}
\newcommand{\bt}{\bar{2}}

\newcommand{\half}{\tfrac{1}{2}}
\newcommand{\sixth}{\tfrac{1}{6}}

\begin{document}

\title{Atomic-Scale Structure of Dislocations Revealed by Scanning
  Tunelling Microscopy and Molecular Dynamics}

\author{J. Christiansen}
\affiliation{CAMP and Department of Physics, Technical University of
  Denmark, DK-2800 Lyngby, Denmark}
\affiliation{Materials Research Department, Ris\o{} National
  Laboratory, DK-4000 Roskilde, Denmark}
\author{K. Morgenstern}
\thanks{Corresponding author}
\email{karina.morgenstern@physik.fu-berlin.de}
\affiliation{Institut f\"ur Experimentalphysik, FB Physik, Freie
  Universit{\"a}t Berlin, Arnimallee 14, D-14195 Berlin, Germany}
\affiliation{\mbox{CAMP and Department of Physics and Astronomy,
  University of Aarhus, DK-8000 Aarhus C, Denmark}}
\author{J. Schi\o{}tz}
\affiliation{CAMP and Department of Physics, Technical University of
  Denmark, DK-2800 Lyngby, Denmark}
\author{K. W. Jacobsen}
\affiliation{CAMP and Department of Physics, Technical University of
  Denmark, DK-2800 Lyngby, Denmark}
\author{K.-F. Braun}
\affiliation{Institut f\"ur Experimentalphysik, FB Physik, Freie
  Universit{\"a}t Berlin, Arnimallee 14, D-14195 Berlin, Germany}
\author{K.-H. Rieder}
\affiliation{Institut f\"ur Experimentalphysik, FB Physik, Freie
  Universit{\"a}t Berlin, Arnimallee 14, D-14195 Berlin, Germany}
\author{E. L\ae{}gsgaard}
\affiliation{\mbox{CAMP and Department of Physics and Astronomy,
  University of Aarhus, DK-8000 Aarhus C, Denmark}}
\author{F. Besenbacher}
\affiliation{\mbox{CAMP and Department of Physics and Astronomy,
  University of Aarhus, DK-8000 Aarhus C, Denmark}}

\date{\today}

\begin{abstract}
  The intersection between dislocations and a Ag(111) surface has been
  studied using an interplay of scanning tunneling microscopy (STM)
  and molecular dynamics (MD).  Whereas the STM provides atomically
  resolved information about the surface structure and Burgers vectors of
  the dislocations, the simulations can be used to determine
  dislocation structure and orientation in the near-surface region.
  In a similar way, the sub-surface structure of other extended
  defects can be studied.  The simulations show dislocations to
  reorient the partials in the surface region leading to an increased
  splitting width at the surface, in agreement with the STM
  observations.  Implications for surface-induced cross slip are
  discussed.
\end{abstract}


\maketitle

Dislocations at surfaces play a major role in many areas of materials
science. Dislocations intersecting surfaces are for example very
important for the control of growth and solidification processes.
Steps terminating at screw dislocations are ideal for continued
crystal growth \cite{HL} and dislocations can strongly modify the
surface stress important for atomic mobility and island nucleation in
growth from the gas phase \cite{surfacestress}.  Dislocation-surface
interactions also play a crucial role in determining the fracture
toughness of a material since the emission and absorption of
dislocations at a crack tip controls the possible blunting of the tip
\cite{CaTh98}.

However, the atomistic understanding of the behavior of dislocations
at surfaces is still very scarce
\cite{Figuera,surfdislocations,marks83,STM-GaAs,expCu,wolf91}.
Bulk
dislocations have in the past been studied extensively with electron
microscopy (EM) and in some cases it has been possible to obtain
atomic-scale information by imaging columns of atoms along a straight
bulk dislocation \cite{Spence99}.  It is not possible to obtain similar resolution with EM for dislocations at surfaces.  STM on
the other hand is the technique of choice to reveal atomic-scale
surface structures such as surface dislocations, but STM is blind with
respect to the region beneath the topmost layer.

In this Letter we show how STM observations of a dislocation
intersecting a surface can be combined with atomistic simulations to
provide detailed atomistic information about the structure of the
dislocation in the near-surface region.  The good agreement between
experimentally observed and simulated surface structures gives
credibility both to the interpretation of the surface structures as
signatures of dislocations and to the calculated structures below the
surface.  The same principles can be used to study the interactions
between surfaces and other extended defects such as grain boundaries.

Consider a bulk dislocation that intersects a surface. If the total
Burgers vector has a nonzero component along the surface normal, a
surface step will end at the dislocation.  If the dislocation is split
into partial dislocations, a step may be seen between the
partial dislocations, even if the total Burgers vector is parallel to
the surface.  From an atomic resolved STM image, the in-plane component of
the Burgers vector can also be determined, allowing a direct
determination of the Burgers vectors of the individual partial dislocations.

The STM cannot, however, determine the edge or screw character of the
dislocations as it gives no information about the line vector of the
dislocation.  Information about the line vector and the sub-surface
structure can be obtained from an interplay with MD simulations.

The STM measurements were performed on a single crystal Ag(111)
surface in two ultra-high vacuum systems equipped with standard
facilities for sample preparation and characterization.  One chamber
houses a fast-scanning, variable-temperature STM \cite{Bes96}, the
other one a low-temperature STM \cite{Meyer96}.  The clean Ag(111)
surface was prepared by several sputtering-annealing cycles.

\begin{figure}[tbp]
   \includegraphics[width=\linewidth]{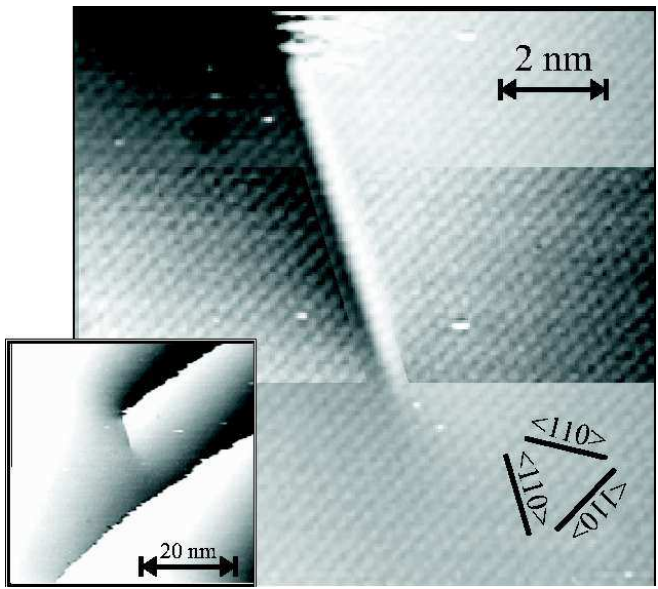}
  \caption{Atomic resolution image of the intersection
    of a bulk dislocation with the Ag(111) surface and its
    surroundings (inset).  The Burgers vector is $\half\langle
    110 \rangle$. For better visibility, the contrast is
    enhanced on a broad stripe in the middle of the image to both the
    left side and the right side of the step, thus no tip change
    occurs.  The horizontal shift in the atomic positions on the two
    sides of the extended dislocation is due to the in-surface
    component of the Burgers vectors of the partial dislocations.
    The STM parameters were $U = -0.22$\,V, $I = 1.4$\,nA,
    and $T = 330$\,K.}
\label{atomic_res}
\end{figure}

Several kinds of dislocations were found in the studied samples.  In a
number of cases, a surface step ends at a dislocation, see
Fig.~\ref{atomic_res}.  Despite a significant amount of surface
diffusion, the structure of the dislocation itself is very stable.  Up
to a step height of about 0.13\,nm, i.e., around 2/3 of the full step
height, the positions of the atoms do not change within half an hour.
Above this step height the step shows the frizziness expected for
Ag(111) at this temperature, as the step is moving while the STM data
are collected
\cite{morgenstern95,poensgen92}.
This inhibits the
determination of the step profile with the same precision as on the
lower parts of the step.

In rare cases dislocations with a Burgers vector in the surface plane
were observed, and in these cases only a minor perturbation of
the surface is seen. In a few cases some features which we interpret as
Lomer-Cottrell locks (sessile dislocations with edge character
\cite{HL}) were seen.

The simulations were performed using a previously determined Effective
Medium Theory (EMT) potential
\cite{orgEMT,modEMT}
for Ag.  The
potential has been fitted to the elastic constants, the vacancy
formation energy, and the intrinsic stacking fault energy in
accordance with experimental and \emph{ab initio} data \cite{newAg}.
The simulations have been done at zero Kelvin by minimizing the energy
of the system, and at 300\,K using Langevin dynamics with a time step
of 5\,fs \cite{AlTi87}.  All simulation cells are rectangular with
(111) surfaces at the top and bottom of the cells, with free boundary
conditions on all surfaces. It is not possible to have
periodic boundary conditions in the directions perpendicular to the
surface due to the net Burgers vector in the simulation cell. Care was
taken to insure that the systems were large enough to prevent unwanted
interactions with the other surfaces of the system.

\begin{figure}[tbp]
\begin{center}
  \includegraphics[width=\linewidth]{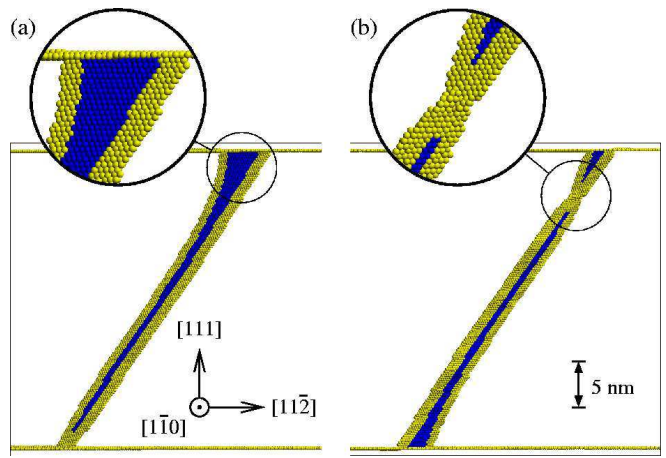}
  \caption{Two configurations of a dislocation with $\boldsymbol{b}=\half
    [110]$, corresponding to two different local minima of the energy.
    The simulated system contains approximately 2 million atoms. The
    atoms in local fcc order and the atoms on the surfaces
    perpendicular to the (111) surfaces of the simulation cell have
    been removed, corresponding to showing 1\,\% of the atoms. The
    atoms in local hcp order are dark while the atoms in the
    dislocation cores are grey. Configuration (a) has the lowest
    energy. Configuration (b) has a constriction where the slip plane
    of the dislocation changes, and this is shown in the blow up.
    Notice how the separation between the partials changes near the
    surfaces.}
  \label{simulconf}
\end{center}
\end{figure}
Fig.~\ref{simulconf} shows two calculated equilibrium configurations
of a dislocation with a Burgers vector of $\boldsymbol{b}=\half
[110]$.  Fig.~\ref{simulconf}(a) corresponds to the lowest energy. The
dislocation can dissociate according to one of the reactions
\begin{equation}
  \half [110] \to \sixth[21\bo] + \sixth[121] \quad
  \text{dissociating on $(1\bo 1)$}\label{eq:splitI}
\end{equation}
\begin{equation}
  \half [110] \to \sixth[12\bo] + \sixth[211] \quad
  \text{dissociating on $(\bo 11)$}.\label{eq:splitII}
\end{equation}
The simulation in Fig.~\ref{simulconf}(a) was set up as two partials
according to reaction \ref{eq:splitI}, with the partials initially
separated by 2\,nm at the surfaces, and the configuration was then
allowed to relax.  The equilibrium configuration has a line vector of
$[110]$ and is thus a screw dislocation, although this does not give
the shortest dislocation length.  The dislocation of
Fig.~\ref{simulconf}(a) was initially set up with this line vector,
but dislocations with other line vectors were seen to rotate to this
orientation, starting at the surfaces.

Linear elasticity theory predicts that a screw
dislocation has lower energy than an edge dislocation and that the
elastic repulsion between two screw dislocations at right angles
vanishes \cite{HL}.  The elastic energy is thus significantly lowered
if the partials rotate away from the bulk orientation to obtain a more
screw-like character.  This process is only possible because the
surfaces break translation symmetry along the dislocation line,
resulting in the changes seen near the surfaces in
Fig.~\ref{simulconf}(a).  Similar effects have been seen in
simulations of a screw dislocation intersecting a Cu(110) surface
\cite{TRcross}.

\begin{figure}[tbp]
  \includegraphics[angle = -90, width=0.8\linewidth]{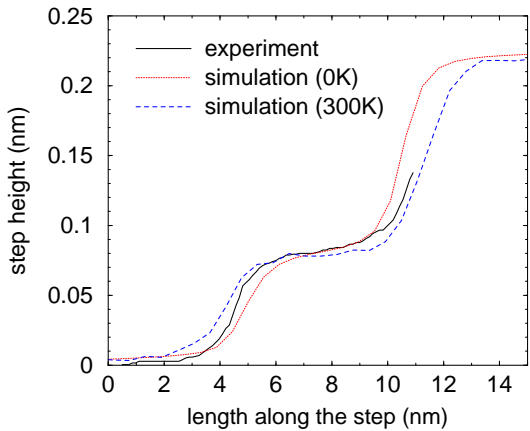}  
  \caption{Step profiles extracted from the STM image in
    Fig.~\protect\ref{atomic_res} (solid line), from the simulation in
  Fig.~\protect\ref{simulconf}(a) ($T=0$\,K, dotted line), and from a
  similar simulation at $T=300$\,K (dashed line).} \label{profiles}
\end{figure}

The surface imprint of the dislocation at the upper surface in
Fig.~\ref{simulconf}(a), where the distance between the partials is
increased, corresponds well with the observed structure in
Fig.~\ref{atomic_res}.  The profile of the surface step ending at the
dislocation is shown in Fig.~\ref{profiles} for both the experiment
and the simulation, and the splitting width as well as the widths of
the individual partials agree well.

The direction of the surface step is not determined by the
dislocation, and it may change if surface diffusion adds or removes
atoms at the step \cite{KMtobepublished}.  In the simulations, the
surface step was placed in the same way as in Fig.~\ref{atomic_res},
i.e.\ adjacent to the partial with the largest out-of-surface component
of the Burgers vector.  If an extra half layer of atoms was added to
the surface, the step would be in the opposite direction and we would
see a high step followed by a low step in Fig.~\ref{profiles}.

Another sub-surface structure is possible for the dislocation, that of
Fig.~\ref{simulconf}(b), obtained when the energy of the perfect
(undissociated) screw dislocation is minimized.  In this case,
dissociation begins at the surfaces according to reaction
\ref{eq:splitI} at the top surface and reaction \ref{eq:splitII} at
the bottom surface, in both cases creating partial dislocations
rotated towards screw character.  The screw-like constriction formed
in Fig.~\ref{simulconf}(b) has a low energy.  In Cu such a
constriction has been shown to have negative energy compared to the
straight dislocation \cite{TRcross}.

\begin{figure}[tp]
\begin{center}
  \includegraphics[width=\linewidth,clip]{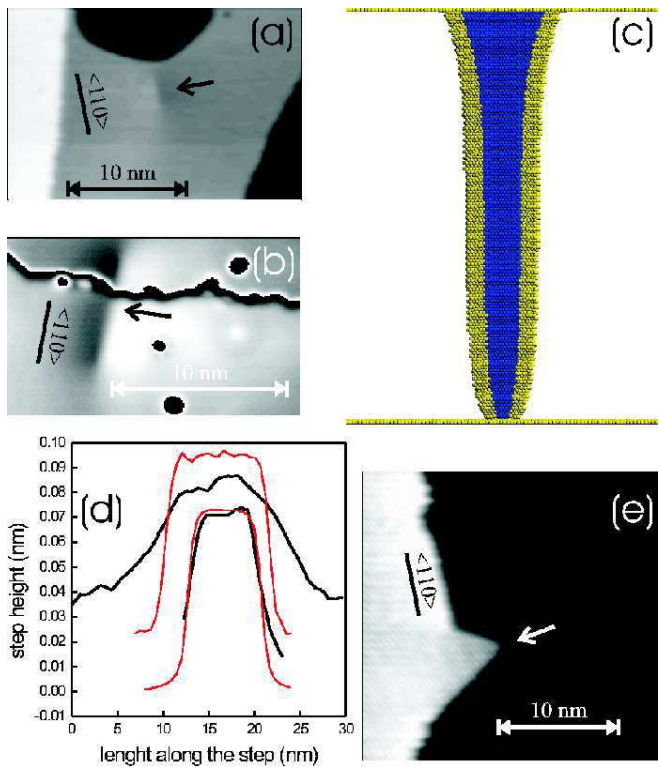}
  \caption{Observations and simulations of dislocations and
    observation of a Lomer-Cottrell lock. In (a) and (b) we show STM
    images of dislocations with Burgers vectors in the surface plane,
    emerging at the arrows. The temperature was $T = 318$\,K and $T =
    7.5$\,K, respectively. The semi-circle in (a) is a vacancy island.
    The dislocation in (b) crosses a surface step.  The image
    constrast has been chosen such that greyscales on the two terraces
    are identical.  The simulated sub-surface structure of such dislocations
    at $T=0$\,K is shown in (c). The step profiles of the dislocations
    (thick lines, (a) at the top, (b) at the bottom) and the simulated
    step profiles at $T= 300$\,K and $T = 0$\,K (thin lines, top and
    bottom, respectively) are shown in (d). The upper curves are
    shifted upwards, and the low-temperature experimental curve has
    been shifted to lie on the simulated curve, since the total height
    could not be determined in the experiment. In (e) we show a
    structure consistent with a Lomer-Cottrell lock pinning a surface
    step.}
  \label{edge}\vspace*{-\baselineskip}
\end{center}
\end{figure}

The energy difference between the two configurations in
Fig.~\ref{simulconf} is 4.5\,eV, favoring the unconstricted
configuration in Fig.~\ref{simulconf}(a).  The surface imprint of the
constricted dislocation in Fig.~\ref{simulconf}(b) is not seen in the
STM.  Notice, however, that TEM studies of dislocations in thin foils
of Cu 10 at.\% Al \cite{Hazzledine-PM, Hazzledine-ICSMA} revealed both
structures in Fig.~\ref{simulconf}.  Also, Fig.~\ref{simulconf}(a)
indicates that a second much lower splitting width of about 1\,nm with
no discernible plateau between the partials should be observed
experimentally, corresponding to the configuration at the lower
surface of Fig.~\ref{simulconf}(a). If the dislocation had been
dissociated on the $(\bo 11)$ plane instead of the $(1\bo 1)$ plane,
the narrow and wide ends would have been swapped and both
configurations are therefore expected at the same surface.  We do,
however, not observe the narrow configuration, except for dislocations
with Burgers vector in the surface plane (see Fig.~\ref{edge}), where
both configurations are seen.

The altered dislocation structure near the surface may influence cross
slip rates of screw dislocations.  Cross slip is the process by which
a screw dislocation changes glide plane and is an important process in
metals in the late stages of work hardening \cite{workhard}.  As the
cross slip process requires  two partial dislocations to first
recombine into a perfect dislocation, the rate depends critically on
the separation of the partials.  The dislocation in figure
\ref{simulconf}(a) will have a much increased probablility of
cross slip near the bottom surface, but not near the top.  The
dislocation in figure \ref{simulconf}(b) has effectively already begun
the cross slip process, which can proceed by moving the constriction
down along the dislocation.

Figs.~\ref{edge}(a) and (b) show dislocations with Burgers vectors in
the surface plane at two different temperatures.  No surface step
ends at the dislocation, but it is nevertheless visible as it splits
into partials that each have a Burgers vector with a component
perpendicular to the surface. The dislocation in Fig.~\ref{edge}(b)
crosses a step, but this is of minor consequence since the elastic
field of a step is small.

In Fig.~\ref{edge}(d) we show step profiles along the dislocations in
Figs.~\ref{edge}(a) and (b) as well as from simulations at $T=300$ K
and $T=0$ K.  The dislocation dissociates according to
the reaction $\half[1 \bo 0] \to \sixth[1 \bt \bo]+\sixth[2 \bo 1]$.
Even though the height of the
experimental curve could not be determined, good agreement is found
between simulations and experiment. The agreement is less good at the
higher temperatures. Part of the experimental broadening is due to
lower resolution of the image itself and especially of the tip. While
the step edge in Fig.~\ref{edge}(b) has an apparent width of 1\,nm
(consistent with a perfect tip), the step edge in Fig.~\ref{edge}(a)
has an apparent width of 2\,nm.

The splitting width of the dislocations, i.e.\ the distance between
the centers of the two partial dislocations, can easily be obtained
from the step profiles.  For the dislocation in Fig.~\ref{atomic_res}
we obtain a splitting with of 6.4 nm, to be compared with 5.5 nm and
7.2 nm in the simulations at 0 K and 300 K, respectively.  The
experimental data is insufficient to reliably extract splitting width
for the dislocations in Fig.~\ref{edge}, the splitting widths in the
simulations are 8.1 nm and 11 nm at 0 K and 300 K, respectively.  All
these values are significantly higher than the bulk splitting widths,
measured to be $d_{\text{s}} \simeq 2.1$\,nm for screw dislocations in
silver, and $d_{\text{e}} \simeq 8.5$\,nm for edge dislocations
\cite{silversplit}.  In our simulations, the splitting widths far from
the surfaces are $d_{\text{s}} \simeq 1.5$ nm and $d_{\text{e}} \simeq
6.6$ nm, in reasonable agreement with the experimental values.

Two mobile dislocations may collide and form a Lomer-Cottrell (LC)
lock which is a sessile edge dislocation that splits on two different
planes. This can, e.g., occur according to the reaction 
\begin{displaymath}
    \half [\bo 01] + \half [01\bo] \to \half [\bo 10]
    \to \sixth [\bo 1\bt] + \sixth [\bo 10] + \sixth [\bo 12]
\end{displaymath}
with the line vector along [110].  A surface structure consistent with
such a dislocation pinning a surface step is seen in
Fig.~\ref{edge}(e). The angle of the sharp v-shape is consistent
with the angle between the splitting planes of a LC lock. The step
height changes by one third at the two Shockley partials, since they 
have a component of the Burgers vector perpendicular to the surface.  
Although there is complete symmetry between the two Shockley partials, 
the lowest energy configuration spontaneously breaks this symmetry. 
Linear elasticity theory predicts a ratio between the splitting
distances of  3.8:1 \cite{LCelastic}.  A lower ratio is clearly seen
in Fig.~\ref{edge}(e), probably from the influence of the surface. 
Unfortunately, the large splitting distances make an atomistic
simulation with a realistic potential prohibitively expensive.

In summary, we have recorded STM images of dislocations intersecting
the Ag(111) surface and measured step profiles of the resulting
surface steps. These STM experiments have been compared to atomic-scale
simulations with good agreement, even on the quantitative level.
We find that this combination of STM and atomic-scale simulations
provide a powerful method for studying the surface-induced structural
changes of crystal defects.

\begin{acknowledgments}
This work was supported by the Danish Research Councils through Grant
No. 5020-00-0012. The Center for Atomic-scale Materials Physics (CAMP) 
is sponsored by the Danish National Research Foundation.
\end{acknowledgments}


\begin{thebibliography}{25}
\expandafter\ifx\csname natexlab\endcsname\relax\def\natexlab#1{#1}\fi
\expandafter\ifx\csname bibnamefont\endcsname\relax
  \def\bibnamefont#1{#1}\fi
\expandafter\ifx\csname bibfnamefont\endcsname\relax
  \def\bibfnamefont#1{#1}\fi
\expandafter\ifx\csname citenamefont\endcsname\relax
  \def\citenamefont#1{#1}\fi
\expandafter\ifx\csname url\endcsname\relax
  \def\url#1{\texttt{#1}}\fi
\expandafter\ifx\csname urlprefix\endcsname\relax\def\urlprefix{URL }\fi
\providecommand{\bibinfo}[2]{#2}
\providecommand{\eprint}[2][]{\url{#2}}

\bibitem[{\citenamefont{Hirth and Lothe}(1968)}]{HL}
\bibinfo{author}{\bibfnamefont{J.~P.} \bibnamefont{Hirth}} \bibnamefont{and}
  \bibinfo{author}{\bibfnamefont{J.}~\bibnamefont{Lothe}},
  \emph{\bibinfo{title}{Theory of Dislocations}}
  (\bibinfo{publisher}{McGraw-Hill}, \bibinfo{year}{1968}).

\bibitem[{\citenamefont{Brune}(1998)}]{surfacestress}
\bibinfo{author}{\bibfnamefont{H.}~\bibnamefont{Brune}},
  \bibinfo{journal}{Surf. Sci. Rep.} \textbf{\bibinfo{volume}{31}},
  \bibinfo{pages}{121} (\bibinfo{year}{1998}).

\bibitem[{\citenamefont{Carlsson and Thomson}(1998)}]{CaTh98}
\bibinfo{author}{\bibfnamefont{A.~E.} \bibnamefont{Carlsson}} \bibnamefont{and}
  \bibinfo{author}{\bibfnamefont{R.}~\bibnamefont{Thomson}},
  \bibinfo{journal}{Solid State Physics} \textbf{\bibinfo{volume}{51}},
  \bibinfo{pages}{233} (\bibinfo{year}{1998}).

\bibitem[{\citenamefont{de~la Figuera~\emph{et al.}}(1998)}]{Figuera}
\bibinfo{author}{\bibfnamefont{J.}~\bibnamefont{de~la Figuera~\emph{et al.}}},
  \bibinfo{journal}{Phys. Rev. B} \textbf{\bibinfo{volume}{58}},
  \bibinfo{pages}{1169} (\bibinfo{year}{1998}).

\bibitem[{\citenamefont{Rodr\'{\i}guez de~la Fuente
  et~al.}(2001)\citenamefont{Rodr\'{\i}guez de~la Fuente, Gonz\'alez, and
  Rojo}}]{surfdislocations}
\bibinfo{author}{\bibfnamefont{O.}~\bibnamefont{Rodr\'{\i}guez de~la Fuente}},
  \bibinfo{author}{\bibfnamefont{M.~A.} \bibnamefont{Gonz\'alez}},
  \bibnamefont{and} \bibinfo{author}{\bibfnamefont{J.~M.} \bibnamefont{Rojo}},
  \bibinfo{journal}{Phys. Rev. B} \textbf{\bibinfo{volume}{63}},
  \bibinfo{pages}{085420} (\bibinfo{year}{2001}).

\bibitem[{\citenamefont{Marks}(1983)}]{marks83}
\bibinfo{author}{\bibfnamefont{L.~D.} \bibnamefont{Marks}},
  \bibinfo{journal}{Phys. Rev. Lett.} \textbf{\bibinfo{volume}{51}},
  \bibinfo{pages}{1000} (\bibinfo{year}{1983}).

\bibitem[{\citenamefont{\mbox{Cox} \emph{et al.}}(1990)}]{STM-GaAs}
\bibinfo{author}{\bibfnamefont{G.}~\bibnamefont{\mbox{Cox} \emph{et al.}}},
  \bibinfo{journal}{Phys. Rev. Lett.} \textbf{\bibinfo{volume}{64}},
  \bibinfo{pages}{2402} (\bibinfo{year}{1990}).

\bibitem[{\citenamefont{\mbox{Samsavar} \emph{et al.}}(1990)}]{expCu}
\bibinfo{author}{\bibfnamefont{A.}~\bibnamefont{\mbox{Samsavar} \emph{et
  al.}}}, \bibinfo{journal}{Phys. Rev. Lett.} \textbf{\bibinfo{volume}{65}},
  \bibinfo{pages}{1607} (\bibinfo{year}{1990}).

\bibitem[{\citenamefont{Wolf and Ibach}(1991)}]{wolf91}
\bibinfo{author}{\bibfnamefont{J.~F.}~\bibnamefont{Wolf}} \bibnamefont{and}
  \bibinfo{author}{\bibfnamefont{H.}~\bibnamefont{Ibach}},
  \bibinfo{journal}{Appl. Phys. A} \textbf{\bibinfo{volume}{52}},
  \bibinfo{pages}{218} (\bibinfo{year}{1991}).

\bibitem[{\citenamefont{Spence}(1999)}]{Spence99}
\bibinfo{author}{\bibfnamefont{J.~C.~H.} \bibnamefont{Spence}},
  \bibinfo{journal}{Mater. Sci. Eng. R} \textbf{\bibinfo{volume}{26}},
  \bibinfo{pages}{1} (\bibinfo{year}{1999}).

\bibitem[{\citenamefont{Besenbacher}(1996)}]{Bes96}
\bibinfo{author}{\bibfnamefont{F.}~\bibnamefont{Besenbacher}},
  \bibinfo{journal}{Rep. Prog. Phys.} \textbf{\bibinfo{volume}{59}},
  \bibinfo{pages}{1737} (\bibinfo{year}{1996}).

\bibitem[{\citenamefont{Meyer}(1996)}]{Meyer96}
\bibinfo{author}{\bibfnamefont{G.}~\bibnamefont{Meyer}}, \bibinfo{journal}{Rev.
  Sci. Instr.} \textbf{\bibinfo{volume}{67}}, \bibinfo{pages}{2960}
  (\bibinfo{year}{1996}).

\bibitem[{\citenamefont{Morgenstern et~al.}(1995)\citenamefont{Morgenstern,
  Rosenfeld, Poelsema, and Comsa}}]{morgenstern95}
\bibinfo{author}{\bibfnamefont{K.}~\bibnamefont{Morgenstern}},
  \bibinfo{author}{\bibfnamefont{G.}~\bibnamefont{Rosenfeld}},
  \bibinfo{author}{\bibfnamefont{B.}~\bibnamefont{Poelsema}}, \bibnamefont{and}
  \bibinfo{author}{\bibfnamefont{G.}~\bibnamefont{Comsa}},
  \bibinfo{journal}{Phys. Rev. Lett.} \textbf{\bibinfo{volume}{74}},
  \bibinfo{pages}{2058} (\bibinfo{year}{1995}).

\bibitem[{\citenamefont{Poensgen et~al.}(1992)\citenamefont{Poensgen, Wolf,
  Frohn, Giesen, and Ibach}}]{poensgen92}
\bibinfo{author}{\bibfnamefont{M.}~\bibnamefont{Poensgen}},
  \bibinfo{author}{\bibfnamefont{J.}~\bibnamefont{Wolf}},
  \bibinfo{author}{\bibfnamefont{J.}~\bibnamefont{Frohn}},
  \bibinfo{author}{\bibfnamefont{M.}~\bibnamefont{Giesen}}, \bibnamefont{and}
  \bibinfo{author}{\bibfnamefont{H.}~\bibnamefont{Ibach}},
  \bibinfo{journal}{Surf. Sci.} \textbf{\bibinfo{volume}{274}},
  \bibinfo{pages}{430} (\bibinfo{year}{1992}).

\bibitem[{\citenamefont{Jacobsen et~al.}(1987)\citenamefont{Jacobsen,
  N{\o}rskov, and Puska}}]{orgEMT}
\bibinfo{author}{\bibfnamefont{K.~W.} \bibnamefont{Jacobsen}},
  \bibinfo{author}{\bibfnamefont{J.~K.} \bibnamefont{N{\o}rskov}},
  \bibnamefont{and} \bibinfo{author}{\bibfnamefont{M.~J.} \bibnamefont{Puska}},
  \bibinfo{journal}{Phys. Rev. B} \textbf{\bibinfo{volume}{35}},
  \bibinfo{pages}{7423} (\bibinfo{year}{1987}).

\bibitem[{\citenamefont{Jacobsen et~al.}(1996)\citenamefont{Jacobsen, Stoltze,
  and N{\o}rskov}}]{modEMT}
\bibinfo{author}{\bibfnamefont{K.~W.} \bibnamefont{Jacobsen}},
  \bibinfo{author}{\bibfnamefont{P.}~\bibnamefont{Stoltze}}, \bibnamefont{and}
  \bibinfo{author}{\bibfnamefont{J.~K.} \bibnamefont{N{\o}rskov}},
  \bibinfo{journal}{Surf. Sci.} \textbf{\bibinfo{volume}{366}},
  \bibinfo{pages}{394} (\bibinfo{year}{1996}).

\bibitem[{\citenamefont{Rasmussen}(2000)}]{newAg}
\bibinfo{author}{\bibfnamefont{T.}~\bibnamefont{Rasmussen}},
  \bibinfo{journal}{Phys. Rev. B} \textbf{\bibinfo{volume}{62}},
  \bibinfo{pages}{12664} (\bibinfo{year}{2000}).

\bibitem[{\citenamefont{Allen and Tildesley}(1987)}]{AlTi87}
\bibinfo{author}{\bibfnamefont{M.~P.} \bibnamefont{Allen}} \bibnamefont{and}
  \bibinfo{author}{\bibfnamefont{D.~J.} \bibnamefont{Tildesley}},
  \emph{\bibinfo{title}{Computer Simulation of Liquids}}
  (\bibinfo{publisher}{Claredon Press}, \bibinfo{address}{Oxford},
  \bibinfo{year}{1987}).

\bibitem[{\citenamefont{Rasmussen et~al.}(1997)\citenamefont{Rasmussen,
  Jacobsen, Leffers, and Pedersen}}]{TRcross}
\bibinfo{author}{\bibfnamefont{T.}~\bibnamefont{Rasmussen}},
  \bibinfo{author}{\bibfnamefont{K.~W.} \bibnamefont{Jacobsen}},
  \bibinfo{author}{\bibfnamefont{T.}~\bibnamefont{Leffers}}, \bibnamefont{and}
  \bibinfo{author}{\bibfnamefont{O.~B.} \bibnamefont{Pedersen}},
  \bibinfo{journal}{Phys. Rev. B} \textbf{\bibinfo{volume}{56}},
  \bibinfo{pages}{2977} (\bibinfo{year}{1997}).

\bibitem[{\citenamefont{\mbox{K. Morgenstern} \emph{et
  al.}}()}]{KMtobepublished}
\bibinfo{author}{\bibnamefont{\mbox{K. Morgenstern} \emph{et al.}}},
  \emph{\bibinfo{title}{\emph{to be published}}}.

\bibitem[{\citenamefont{Hazzledine et~al.}(1975)\citenamefont{Hazzledine,
  Karnthaler, and Wintner}}]{Hazzledine-PM}
\bibinfo{author}{\bibfnamefont{P.~M.} \bibnamefont{Hazzledine}},
  \bibinfo{author}{\bibfnamefont{H.~P.} \bibnamefont{Karnthaler}},
  \bibnamefont{and} \bibinfo{author}{\bibfnamefont{E.}~\bibnamefont{Wintner}},
  \bibinfo{journal}{Phil. Mag.} \textbf{\bibinfo{volume}{32}},
  \bibinfo{pages}{81} (\bibinfo{year}{1975}).

\bibitem[{\citenamefont{Wintner et~al.}(1976)\citenamefont{Wintner, Karnthaler,
  and Hazzledine}}]{Hazzledine-ICSMA}
\bibinfo{author}{\bibfnamefont{E.}~\bibnamefont{Wintner}},
  \bibinfo{author}{\bibfnamefont{H.~P.} \bibnamefont{Karnthaler}},
  \bibnamefont{and} \bibinfo{author}{\bibfnamefont{P.~M.}
  \bibnamefont{Hazzledine}}, in \emph{\bibinfo{booktitle}{Proceedings of the
  4th International Conference on the Strength of Metals and Alloys}}
  (\bibinfo{organization}{Laboratoire de Physique du Solide, Nancy},
  \bibinfo{year}{1976}), vol.~\bibinfo{volume}{2}, p. \bibinfo{pages}{927}.

\bibitem[{\citenamefont{Seeger et~al.}(1959)\citenamefont{Seeger, Berner, and
  Wolf}}]{workhard}
\bibinfo{author}{\bibfnamefont{A.}~\bibnamefont{Seeger}},
  \bibinfo{author}{\bibfnamefont{R.}~\bibnamefont{Berner}}, \bibnamefont{and}
  \bibinfo{author}{\bibfnamefont{H.}~\bibnamefont{Wolf}}, \bibinfo{journal}{Z.
  Physik} \textbf{\bibinfo{volume}{155}}, \bibinfo{pages}{247}
  (\bibinfo{year}{1959}).

\bibitem[{\citenamefont{Cockayne et~al.}(1971)\citenamefont{Cockayne, Jenkins,
  and Ray}}]{silversplit}
\bibinfo{author}{\bibfnamefont{D.~J.~H.} \bibnamefont{Cockayne}},
  \bibinfo{author}{\bibfnamefont{M.~L.} \bibnamefont{Jenkins}},
  \bibnamefont{and} \bibinfo{author}{\bibfnamefont{I.~L.~F.}
  \bibnamefont{Ray}}, \bibinfo{journal}{Phil. Mag.}
  \textbf{\bibinfo{volume}{24}}, \bibinfo{pages}{1383} (\bibinfo{year}{1971}).

\bibitem[{\citenamefont{Bonneville and Douin}(1990)}]{LCelastic}
\bibinfo{author}{\bibfnamefont{J.}~\bibnamefont{Bonneville}} \bibnamefont{and}
  \bibinfo{author}{\bibfnamefont{J.}~\bibnamefont{Douin}},
  \bibinfo{journal}{Phil. Mag. Lett.} \textbf{\bibinfo{volume}{62}},
  \bibinfo{pages}{247} (\bibinfo{year}{1990}).

\end{thebibliography}

\newcommand{\noopsort}[1]{} \newcommand{\printfirst}[2]{#1}
  \newcommand{\singleletter}[1]{#1} \newcommand{\switchargs}[2]{#2#1}



\end{document}